\documentclass[a4paper,11pt]{article}
\usepackage{pos}

\usepackage{units}
\graphicspath{{figures/}} 

\title{Science verification of the new FlashCam-based camera in the 28\,m telescope of H.E.S.S.}
\ShortTitle{Science verification of the new FlashCam-based camera in the 28\,m telescope of H.E.S.S.}

\author*[a]{Gerd P\"uhlhofer}
\author[b]{Konrad Bernl\"ohr}
\author[a]{Baiyang Bi}
\author[b]{German Hermann}
\author[b]{Jim Hinton}
\author[c]{Ira Jung-Richardt}
\author[a]{Fabian Leuschner}
\author[b]{Vincent Marandon}
\author[d]{Alison Mitchell}
\author[c]{Lars Mohrmann}
\author[e]{Daniel Parsons}
\author[b]{Simon Sailer}
\author[a]{Heiko Salzmann}
\author[b]{Simon Steinmassl}
\author[b]{Felix Werner}

\affiliation[a]{Institut f\"ur Astronomie und Astrophysik, Eberhard Karls Universit\"at T\"ubingen, Sand 1, 72076 T\"ubingen, Germany}
\affiliation[b]{Max-Planck-Institut f\"ur Kernphysik, PO Box 103980, 69029 Heidelberg, Germany}
\affiliation[c]{Physikalisches Institut, Friedrich-Alexander Universit\"at Erlangen-N\"urnberg, Erwin-Rommel-Str. 1, 91058 Erlangen, Germany}
\affiliation[d]{ETH Z\"urich, Otto-Stern-Weg 5, 8093 Z\"urich, Switzerland}
\affiliation[e]{DESY, D-15738 Zeuthen, Germany}

\forColl{H.E.S.S.} 

\emailAdd{Gerd.Puehlhofer@astro.uni-tuebingen.de}

\abstract{In October 2019 the central 28\,m telescope of the H.E.S.S. experiment has been upgraded with a new camera. The camera is based on the FlashCam design which has been developed in view of a possible future implementation in the medium-sized telescopes of the Cherenkov Telescope Array (CTA). We report here on the results of the science verification program that has been performed after commissioning of the new camera, to show that the camera and software pipelines are working up to expectations.}

\FullConference{37$^{\rm{th}}$ International Cosmic Ray Conference (ICRC 2021)\\
		July 12th -- 23rd, 2021\\
		Online -- Berlin, Germany}

\begin{document}
\maketitle

\section{Introduction}

\begin{figure}
\begin{center}
\includegraphics[width=0.7\textwidth]{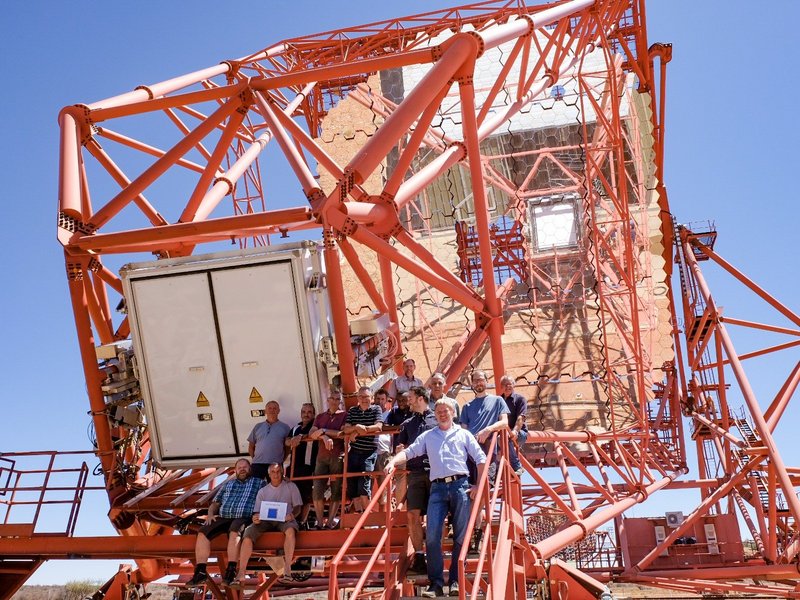} 
\end{center}
\vspace{-0.3cm}
\caption{CT5-FlashCam after successful installation into H.E.S.S.-CT5. Image credit: C. F\"ohr (MPIK).}
\label{fig:telescope}
\end{figure}

The H.E.S.S. collaboration runs a system of five Imaging Atmospheric Cherenkov Telescopes (IACTs) in the Khomas Highland of Namibia. Four 12\,m telescopes (CT1-4) have been running since 2003\footnote{Noteworthy is an upgrade of the CT1-4 camera electronics in 2017.}. A fifth telescope (CT5) was added in 2012, significantly reducing the energy threshold of the system down to \unit[$\sim$30]{GeV} with its \unit[28]{m} dish. In 2019, the focal plane instrument of CT5 was replaced with a new camera (Fig.\,\ref{fig:telescope}). The goal of the exchange was mainly to improve telescope uptime and data quality, compared to the status of the previous camera which had suffered from aging symptoms after several years of operation in the field. 
The replacement camera is an advanced prototype of the FlashCam camera (``CT5-FlashCam''). The photomultiplier-tube-(PMT)-based FlashCam camera type has been developed by a German-Swiss-Austrian team as a possible camera for the Cherenkov Telescope Array (CTA) Medium-Sized Telescopes (MST) \cite{FlashCam2013,FlashCam2019,MST2017}. It is based on FlashCam technology, the first fully-digital trigger and readout system for IACT cameras \cite{FlashCam2008}. 

CT5-FlashCam uses a mix of Hamamatsu R12992-100 PMTs with 7 dynodes and Hamamatsu R11920-100 PMTs with 8 dynodes\footnote{Both PMT types have been under evaluation for use in CTA. Either provides adequate performance for FlashCam.} to detect the incoming photons. 1758 PMTs are arranged in a hexagonal pattern with a 50 mm pitch, with hollow light guides in front to optimize light collection and restrict the incoming photon angles. The overall hexagonal shape of the photon detection plane (PDP) has an area of \unit[3.81]{m$^2$}, 
the diameter of the field of view (FoV) is $3.4^\circ$ in CT5 ($7.7^\circ$ in CTA-MSTs)\footnote{The diameter of the FoV is defined here as twice the average distance of all edge pixels from the FoV center.}. 
The 588 PDP modules (each hosting 12 PMTs, high voltage supply, and pre-amplifiers) deliver the analog signals (one per pixel) to the FlashCam readout electronics. The preamplifier changes from a linear to a quasi-logarithmic behavior above $\sim$$250$\,p.e., yielding a dynamic range up to $\sim$$3000$\,p.e.\ with a single gain and readout channel per pixel. 

The FlashCam readout system is arranged in crates and racks towards the rear side of the camera. 12-bit flash analog-to-digital converters (FADC) modules (each serving 24 pixels) digitize the incoming signals with \unit[250]{MHz} FADCs and process the data with one Spartan 6 FPGA per module. Trigger processing -- in communication with 12 trigger modules (each serving up to 8 FADC modules) and one master module -- as well as data buffering and further processing is fully digital. Triggered event data are transmitted via an ethernet-based protocol to a camera server, which is located at the central array computing room. Camera synchronisation (for system trigger and event timestamping) is provided by a White Rabbit system\footnote{\href{https://ohwr.org/project/white-rabbit/wikis/home}{https://ohwr.org/project/white-rabbit/wikis/home}} via a single fiber. The camera is fully sealed, with slight overpressure inside to prevent dust from entering. By a combination of air ventilation inside the camera and a water chiller system, the heat is removed from the camera.

To limit the data volume to a level that is acceptable for the H.E.S.S.\ DAQ system, storage, and outbound bandwidth, two main changes were made compared to the (anticipated) operation in an array of CTA-MSTs. First, the trigger threshold was adjusted to limit the fraction of accidental triggers due to NSB fluctuations to O(1\%) or less, resulting in monoscopic trigger rates of around 2.5 kHz for the standard thresholds of 69\,p.e.\ in a 9-pixel patch.\footnote{Pixel patch sums are computed in sliding windows, the smallest non-overlapping unit being a 3-pixel patch. To reduce the influence of bright patches in the FoV, pixels with NSB rates exceeding 1.1\,GHz p.e.\ do not contribute to the overlapping 9-pixel patch sums.} Second, to reduce the overall data volume, the pixel waveforms are analysed on the camera server and reduced to pulse shape and timing parameters before transmitting them to the H.E.S.S.\ DAQ.

\section{On the mechanical aspects of the camera exchange}

To mount the FlashCam camera, which was designed for CTA MSTs with corresponding mechanical interface definitions, into the CT5 camera clamping system, a mechanical interface frame had to be constructed. In order to minimize a potential change of the telescope dish bending behavior, the weight of the interface frame was adjusted such that the net weight load on the telescope masts was not changed by the camera exchange. Counterweights were therefore not modified. 

The CT5 camera clamping system features a motorized focusing system, by which the camera can be adjusted along the optical axis. Regularly, this focus system has two settings. For regular observations, the camera's entrance plane (defined by the front of the light guides)
is placed into the telescope's image plane that corresponds to the object being at 15\,km distance\footnote{The typical shower height is in that distance scale.}. The other position places a diffusively reflecting screen in the telescope's focal plane; the screen can on demand be placed in front of the camera. This allows the (on-axis) point spread function (PSF) of the telescope 
to be measured by tracking a star and taking images of the star's spot on the screen, using an optical CCD camera (the ``Lid CCD'') which is located in the telescope dish. After the Cherenkov camera exchange, the focus settings were readjusted, taking the different distances between camera entrance plane and screen of the different Cherenkov camera models into account. 
A readjustment of the CT5 mirror facets was -- as anticipated -- not necessary\footnote{The CT5 dish steel structure is -- despite its size -- stiff enough so that a dynamic readjustment of mirrors depending on telescope elevation is not necessary. Only one mirror alignment state exists for the telescope.}, as was verified through measurements of the telescope's identical PSF before and after camera exchange.

\section{Updated calibration procedures and reconstruction software}

CT5-FlashCam's pixel conversion functions from the number of PMT photoelectrons (p.e.) 
to the measured electronic signal 
have been extensively investigated and calibrated with a laser light source in the lab, before shipment to the H.E.S.S.\ site \cite{FlashCam2017}. Because of the non-linear behavior of the signal chain at signal amplitudes $\gtrsim 250$\,p.e., this is not just a single conversion factor per pixel but a model of the calibration curve. An end-to-end flatfield to level out the slightly different -- and over time changing -- responses\footnote{essentially the product of light guide reflectivity, PMT quantum efficiency, and PMT gain} of the individual pixels to the incoming light is derived using the same flat-fielding unit that had already been in use for the original CT5 camera. The device is an LED flasher located in the telescope dish center which illuminates the camera homogeneously with flashes corresponding to $\sim$130\,p.e.\ per pixel. 
Data from this flat-fielding unit are also used to monitor the PMT gain change (resulting in an overall flattening of the conversion function) and to calibrate the conversion factor (usually called single photoelectron (SPE) response) in the linear regime. The values are derived based on the statistical properties of the light pulses from the CT5 flat-fielding unit, assuming a Poisson distribution of the injected charge and constant excess noise of the PMTs. No special calibration device -- e.g.\ to provide low light illumination to produce SPE spectra -- is needed for that purpose. 
More details can be found in \cite{FlashCam2021}.

Most relevant for correct shower energy reconstruction is the absolute intensity calibration of the camera, i.e.\ the electronic signal size as a function of the incoming Cherenkov photon density on the telescope's dish. This entire -- optical plus electronic -- telescope throughput is calibrated by means of (gain-calibrated and flatfielded) ring images in the camera induced by local\footnote{i.e.\ hitting directly or near the telescope dish} muons, which are released in hadronic air showers. This calibration factor -- one number per entire camera -- is derived continuously (normally per individual observation run of 28 min duration) from the regular data stream. The same method is applied for all H.E.S.S.\ telescopes. 
The corresponding analysis software which identifies muon rings and reconstructs their image parameters, in observational data as well as in corresponding shower and detector simulations, was adapted for the new camera.

Finally, the calibration of the telescope pointing was also adapted to the new camera. The pointing of H.E.S.S.\ telescopes is corrected in software offline with 
a pointing model, to eliminate the slight mispointing during observations which are of the order of (few) arcminutes. 
The mispointing is caused by slight imperfections of the telescope alignment and structural bending under different elevation angles. 
The pointing model is regularly -- typically a few times per year -- calibrated by tracking stars that are homogeneously distributed in telescope altitude and azimuth.
The position of the star's spot on the reflective screen is measured, using the LidCCD camera, relative to a set of optical reference LEDs, which are located at the edge of the Cherenkov camera. 
The pointing calibration software was adapted to the new Cherenkov camera essentially by changing the nominal reference LED positions in the reconstruction software. For CT5 standalone (``mono'') reconstruction, the expected systematic error for a source location continues to be around 25 arcsec (68\,\% c.l.), as estimated from the pointing calibration procedure.

To be able to analyze CT5-FlashCam data, standard H.E.S.S.\ software pipelines were adapted to the new camera. The full camera hardware simulation, which was developed already for CTA performance simulations, was incorporated in the CT5 simulation framework. Image cleaning algorithms (mainly tail-cut thresholds in units of p.e.\footnote{Tail-cuts basically remove non-contiguous low-intensity pixels from the shower images.}) were adapted to reflect the higher quantum efficiencies of the new camera's PMTs. The new pixel geometry was incorporated in simulations and reconstruction software. 
Reconstruction algorithms, like multivariate analyses for background rejection and shower template analyses for $\gamma$-ray reconstruction, were trained on the new simulations. Background acceptance templates across the FoV were initially trained on cosmic ray simulations, and replaced by real background data once available from the instrument in the telescope.

\section{Science verification program}

After commissioning of the new camera in CT5, 
a dedicated observation program was performed to verify that the camera performs up to expectations. Goals of this program were to:

\noindent $\bullet$ Obtain a large number of $\gamma$-ray events from strong (point-like) sources of known spectral shape, to verify that simulations match actual observational data, e.g.\ image parameter distributions.\\
$\bullet$ Verify that the pointing corrections work as expected after the camera exchange, by comparing reconstructed positions of known point-like $\gamma$-ray sources to catalog values. \\
$\bullet$ Compare the measured $\gamma$-ray point spread function (using point sources) to simulations (which match the observations' zenith angle distribution and the target's energy spectrum, respectively), to verify simulations and the derived $\gamma$-ray angular cut efficiencies.\\
$\bullet$ Check the flatness of the background distribution in the FoV, to verify the homogeneity of the camera trigger, the robustness of background rejection algorithms, and the validity of acceptance models for the residual background which are used for background estimates at the source location.\\
$\bullet$ Reconstruct the energy spectrum of a reference $\gamma$-ray point source and compare to expectations, to verify that the optical throughput calibration is correct, that the simulated $\gamma$-ray acceptance of the applied cuts is correct, and that the simulated camera trigger behavior leads to a correctly computed instrument's effective area.\\
$\bullet$ Produce a phasogram of a pulsed reference source with known period and compare to expectations, to verify that the event time stamping works.

To that end, observations were conducted on the Crab Nebula (13.8\,hrs, mean zenith angle 48$^\circ$, goals 1.-5.), PKS\,2155-304 (45.2\,hrs, 24$^\circ$, goals 1.-4.), and the Vela pulsar (10.1\,hrs, 30$^\circ$, goal 6.). The PKS\,2155-304 data set used here also comprises observations taken under the regular observing program that started after the science verification program. In addition, data from PKS\,0903-57 (13.7\,hrs, 39$^\circ$, goals 2.-4.) were used, which were obtained under target of opportunity observations triggered by an outburst of the source detected with \textit{Fermi}-LAT, that led to the discovery of a strong TeV $\gamma$-ray flux from the source.

\section{Results}

In the following, some of the key results of the science verification program of CT5-FlashCam are illustrated. For the results presented in this paper, a preliminary release of the HAP software, which incorporates the CT5-FlashCam-related software and configuration updates as discussed above, was used. HAP is one of the H.E.S.S. standard pipeline software packages. All results were derived with CT5 mono reconstruction. Except for the Vela pulsar analysis, the reconstruction was performed applying a conservative image analysis threshold of 250 p.e.

\begin{figure}
\begin{center}
\includegraphics[width=0.48\textwidth]{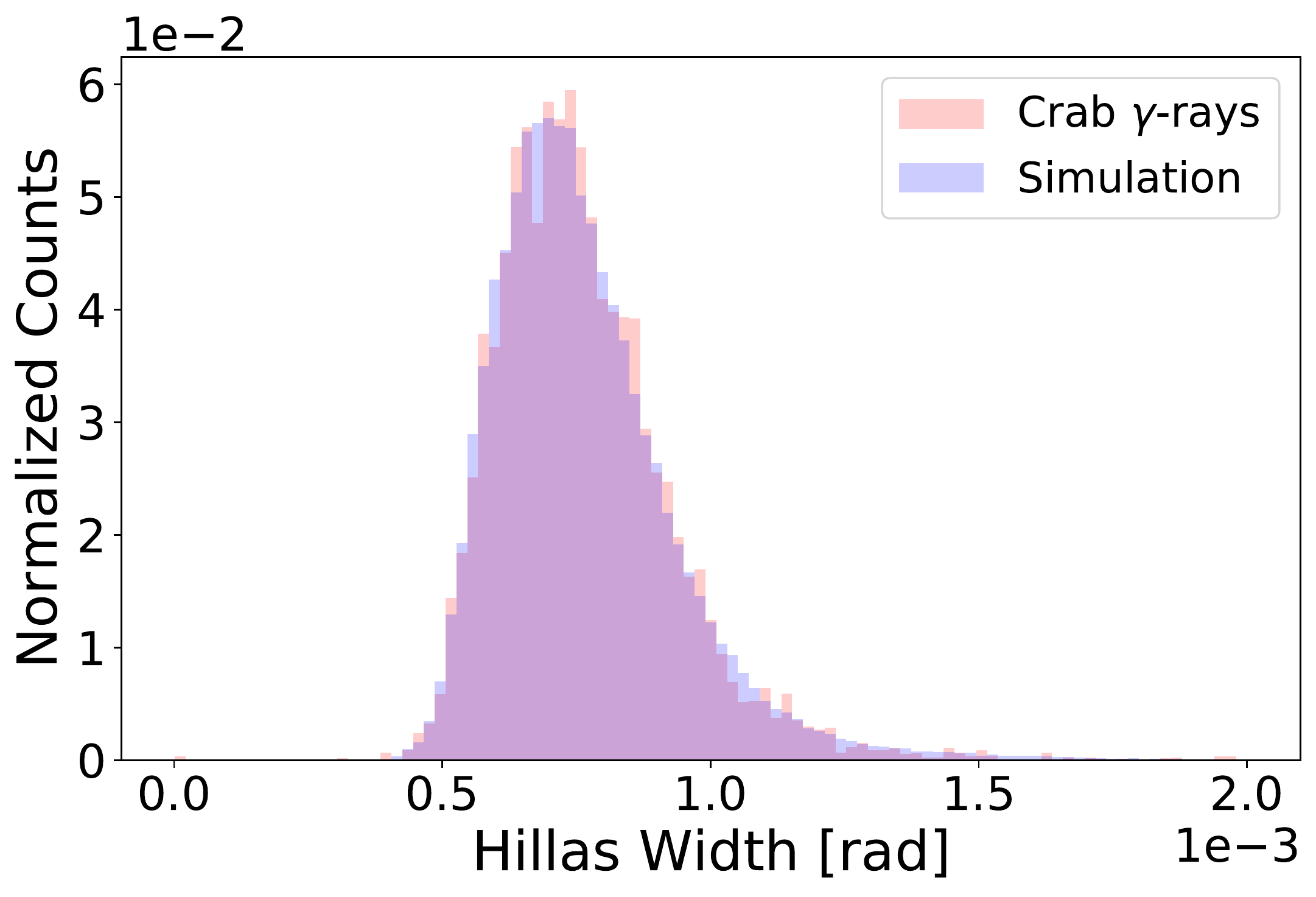}
\hfill
\includegraphics[width=0.48\textwidth]{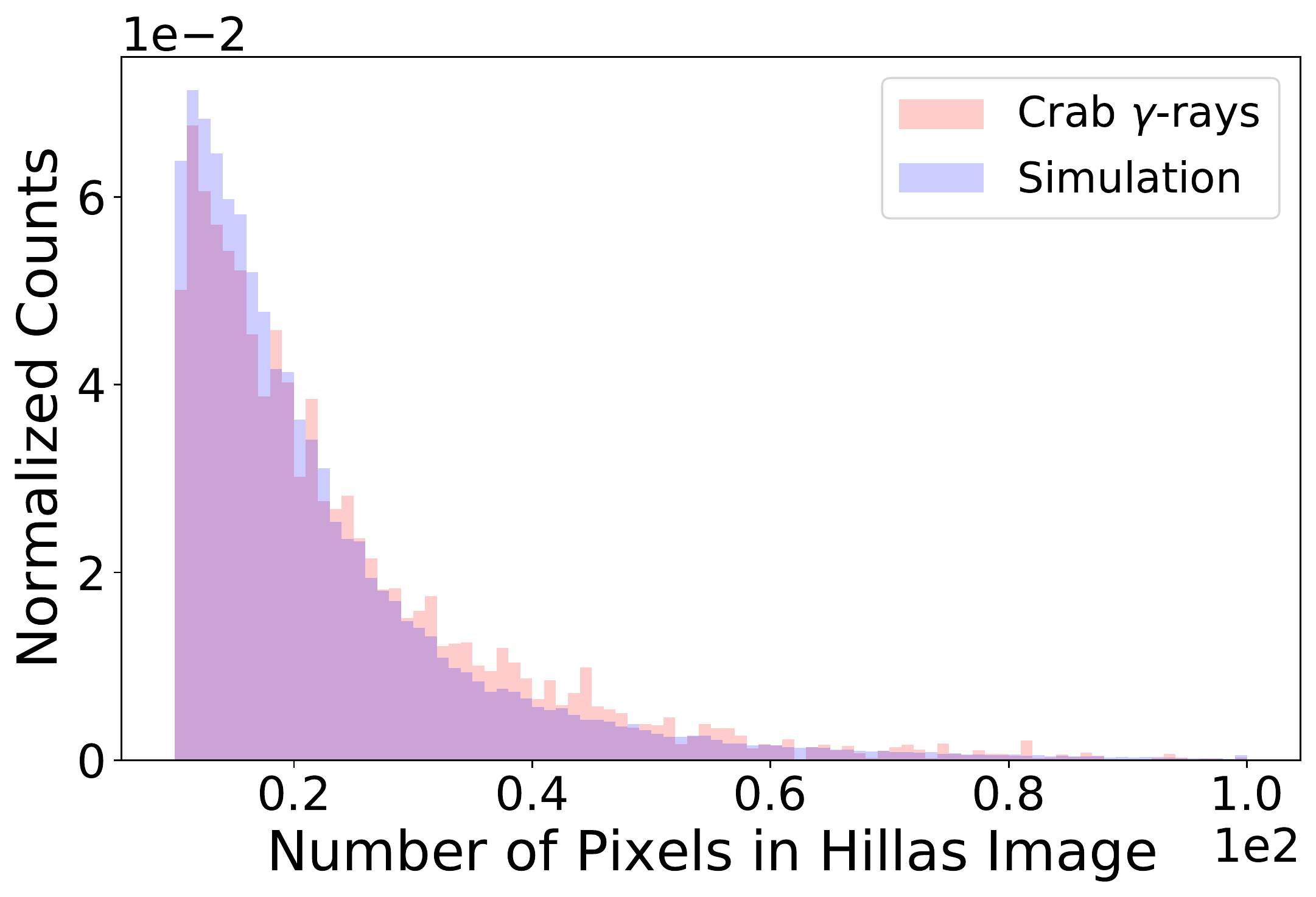}
\end{center}
\vspace{-0.5cm}
\caption{Distributions of the Hillas $\gamma$-ray image width (left panel) and 
number of pixels per $\gamma$-ray image (right panel) derived from Crab Nebula observations (pink-shaded histograms), compared to simulations of matching observation conditions (blue histograms).
}
\label{fig:hillaswidth}
\end{figure}

\begin{figure}
\begin{center}
\includegraphics[trim=30 0 30 0,clip,width=0.48\textwidth]{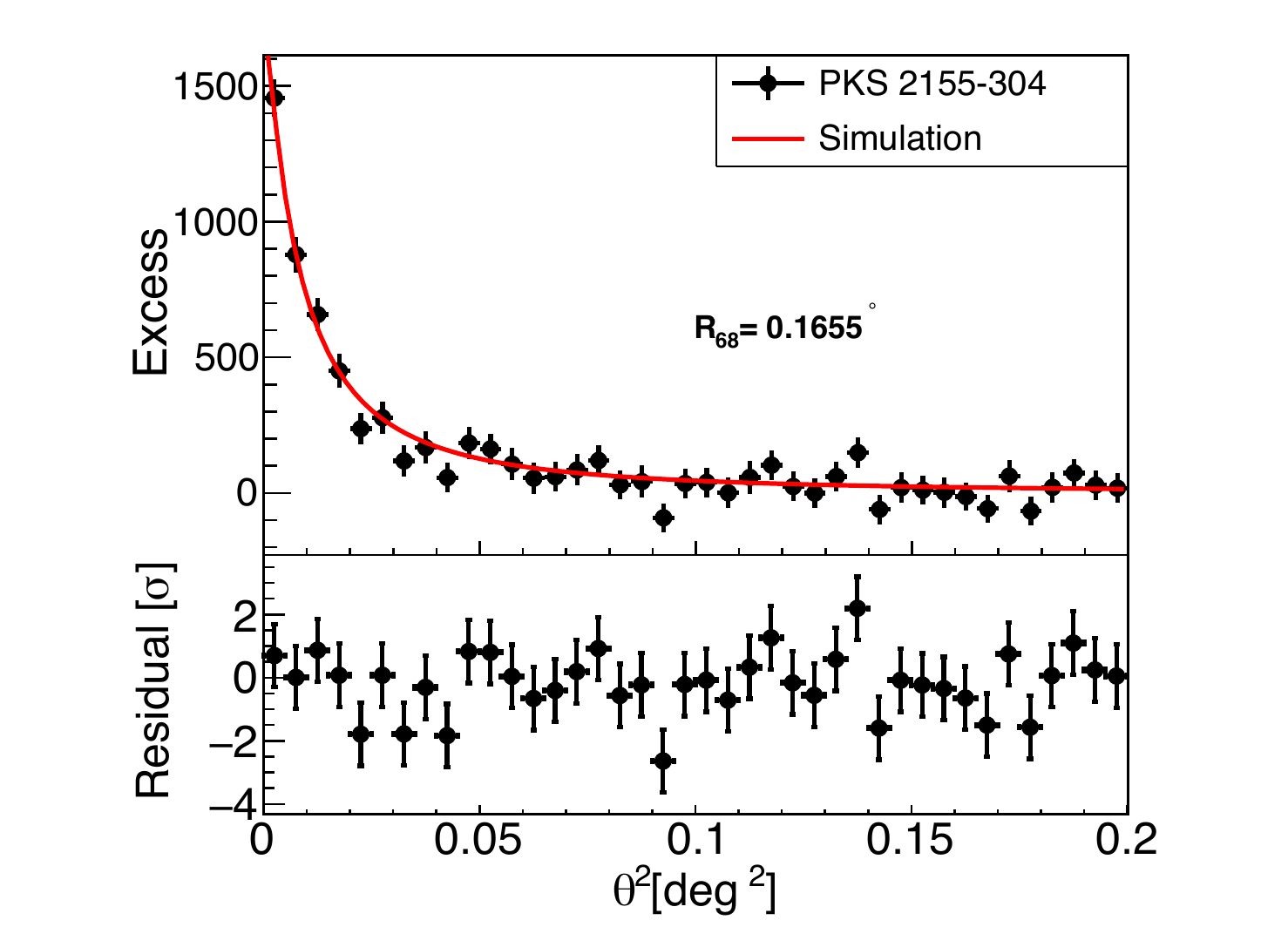}
\hfill
\includegraphics[trim=30 0 30 0,clip,width=0.48\textwidth]{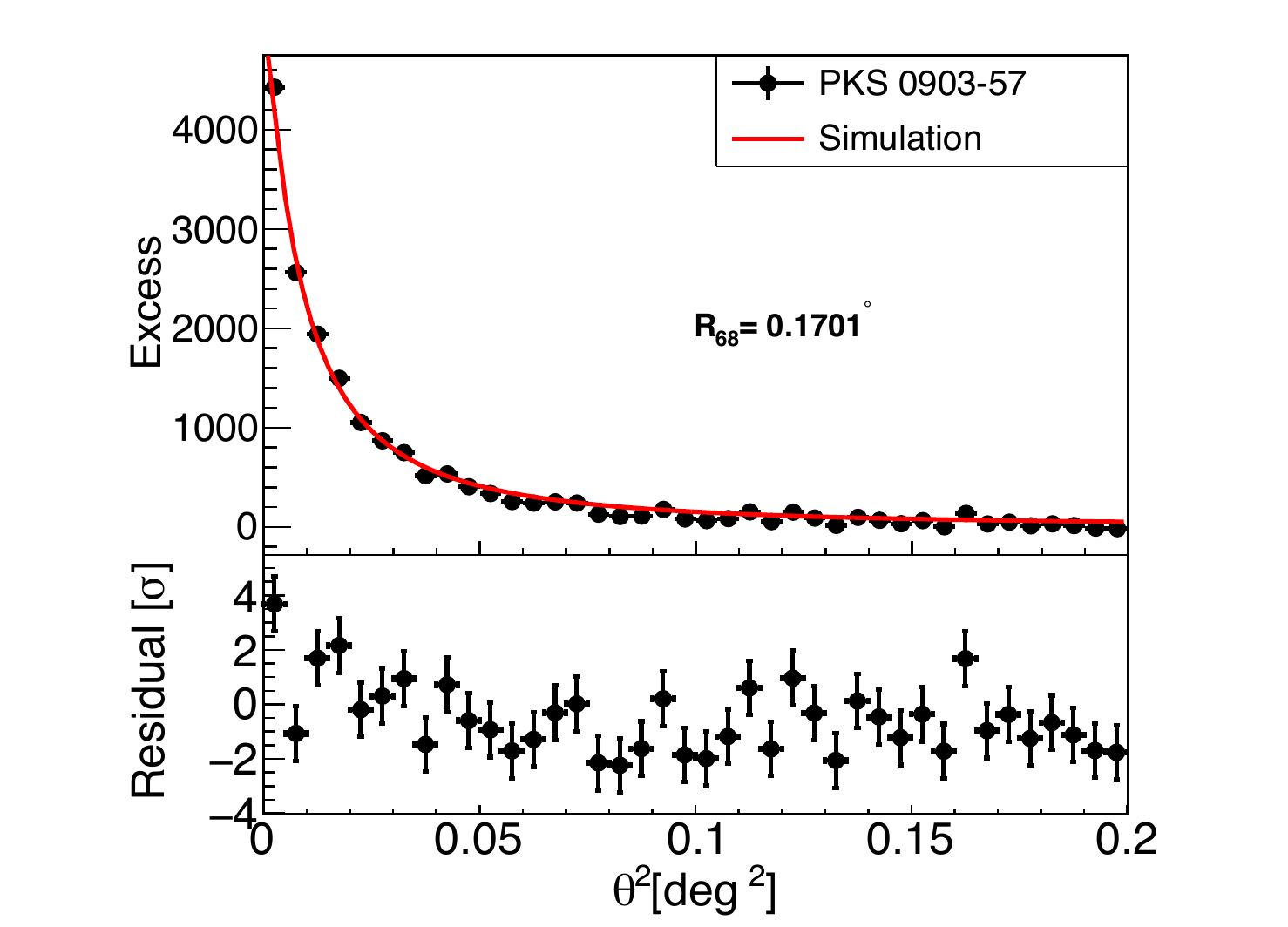}
\end{center}
\vspace{-0.5cm}
\caption{Radial profiles of $\gamma$-rays obtained from PKS\,2155-304 (left panel) and PKS\,0903-57 (right panel), compared to simulations of matching observation conditions. 
}
\label{fig:psf}
\end{figure}

\begin{figure}
\begin{center}
\includegraphics[trim=0 0 0 70,clip,width=0.33\textwidth]{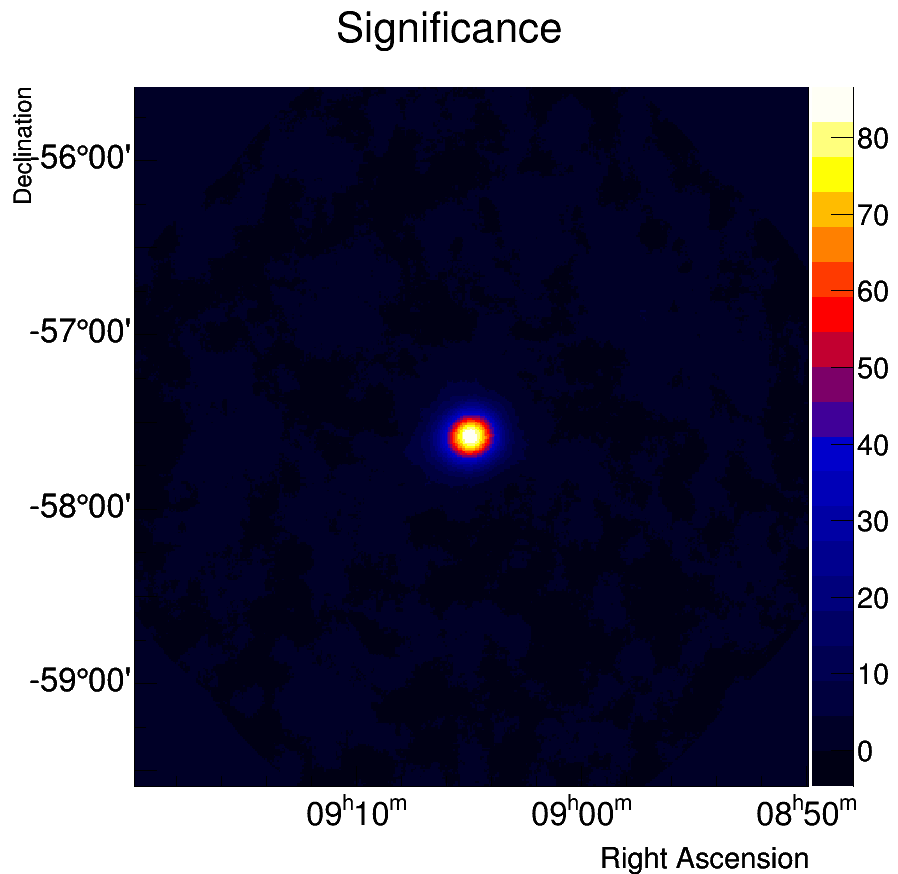}
\hfill
\includegraphics[trim=0 0 0 70,clip,width=0.33\textwidth]{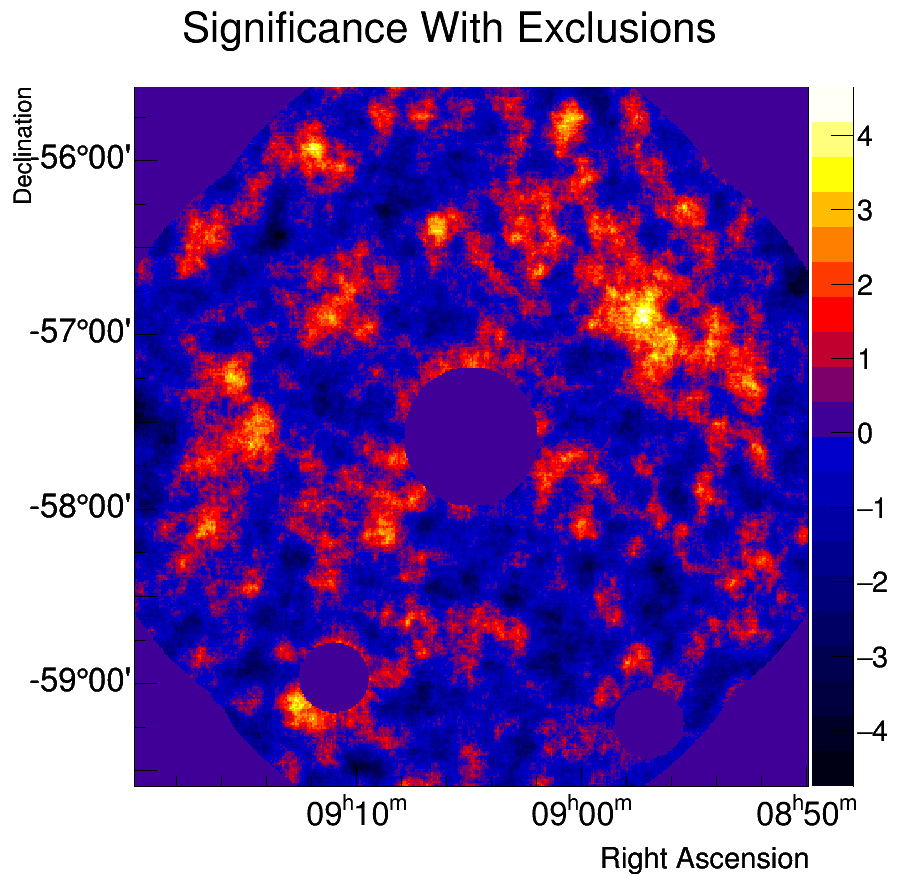}
\hfill
\includegraphics[trim=30 20 30 30,clip,width=0.30\textwidth]{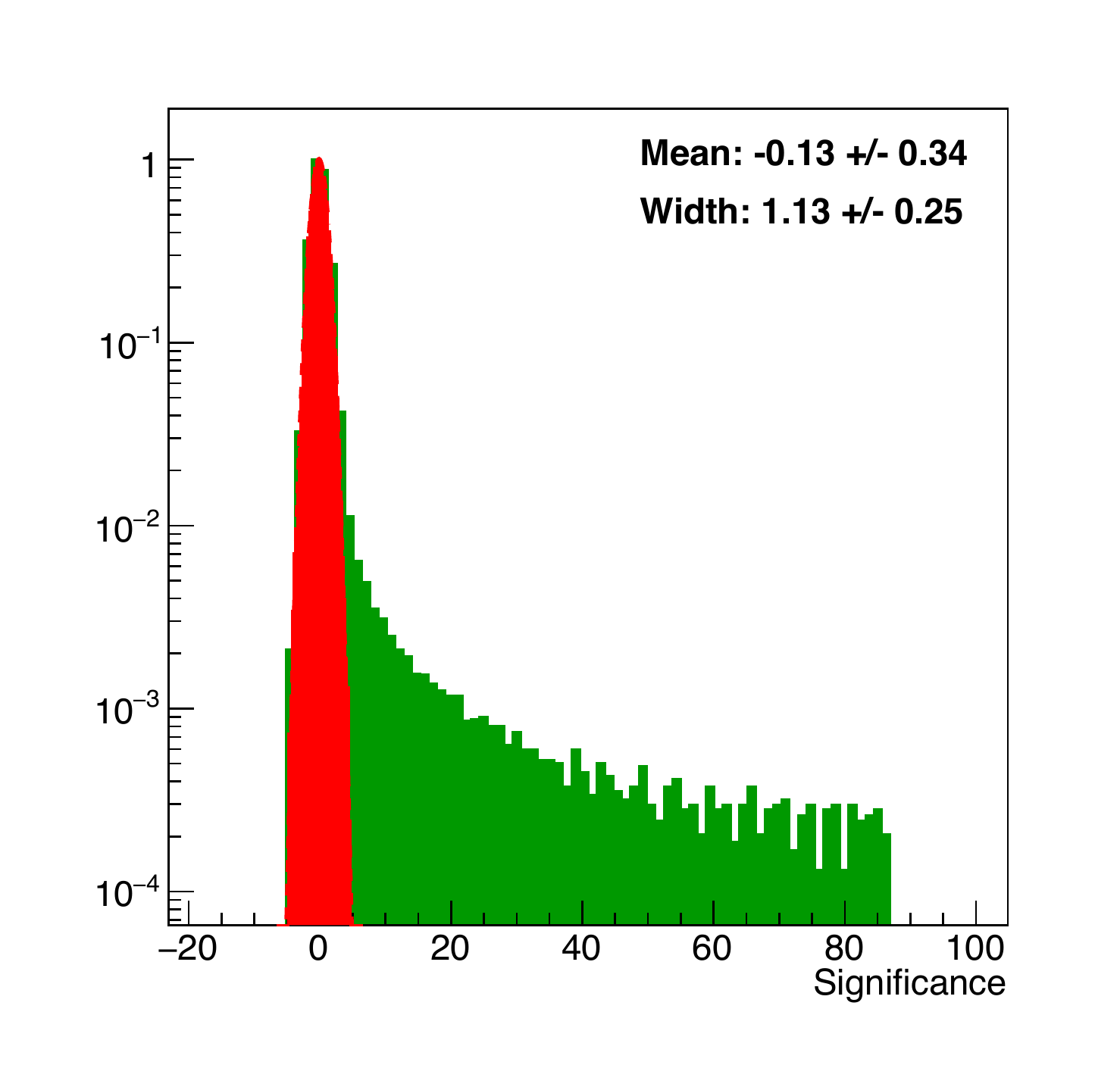}
\end{center}
\vspace{-0.5cm}
\caption{The field of view of PKS\,0903-57 observed with CT5-FlashCam. Left: Significance map of the entire field. Middle: The same significance map, after masking the $\gamma$-ray source itself and two bright stars in the South which cause some analysis artefacts. Right: Significance distribution of the entire field (green histogram), overplotted with a Gaussian fit (red) to the significances of the masked field. 
}
\label{fig:fov}
\end{figure}

Fig.\,\ref{fig:hillaswidth} shows the comparison of the measured and simulated distributions of the Hillas width and number-of-pixels parameters, derived from Crab Nebula data, demonstrating good agreement.

Fig.\,\ref{fig:psf} shows the radial profile of reconstructed $\gamma$-ray events versus the position of PKS\,2155-304 and PKS\,0903-57, respectively, as a function of distance-squared (``theta-squared'' distribution), in comparison to expected profiles (normalized to the measured distribution). These expected profiles represent instrument simulations for matching observation conditions, to which a King's profile\footnote{$K(\theta^2) = K_0  \left(1 - \frac{1}{\gamma}\right) \cdot  \left(1 + \frac{1}{2 \gamma} \cdot \frac{\theta^2}{\sigma^2}\right)^{-\gamma}$, with normalization $K_0$ and $\gamma, \sigma$ describing the shape, for references and naming see \href{https://fermi.gsfc.nasa.gov/ssc/data/analysis/documentation/Cicerone/Cicerone_LAT_IRFs/IRF_PSF.html}{https://fermi.gsfc.nasa.gov/ssc/data/analysis/documentation/Cicerone/Cicerone\_LAT\_IRFs/IRF\_PSF.html}}  was fitted. For both sources -- as also for the Crab Nebula -- the reconstructed source positions were verified to be consistent with expectations from pointing systematics, within 95\,\% c.l.

Fig.\,\ref{fig:fov} illustrates the properties of the sky map of PKS\,0903-57. The significance sky maps 
are computed binwise with an oversampling circle of radius $0.1^\circ$ for on-source events, and rings around each bin position with inner radius of $0.5^\circ$ and outer radius of $0.6^\circ$ for collecting background events. The fit to the background significance distribution (red Gaussian in the right panel of Fig.\,\ref{fig:fov}) is compatible with a mean of 0 and a sigma of 1, as expected for pure statistical noise fluctuations. 

\begin{figure}
\begin{center}
\includegraphics[width=0.78\textwidth]{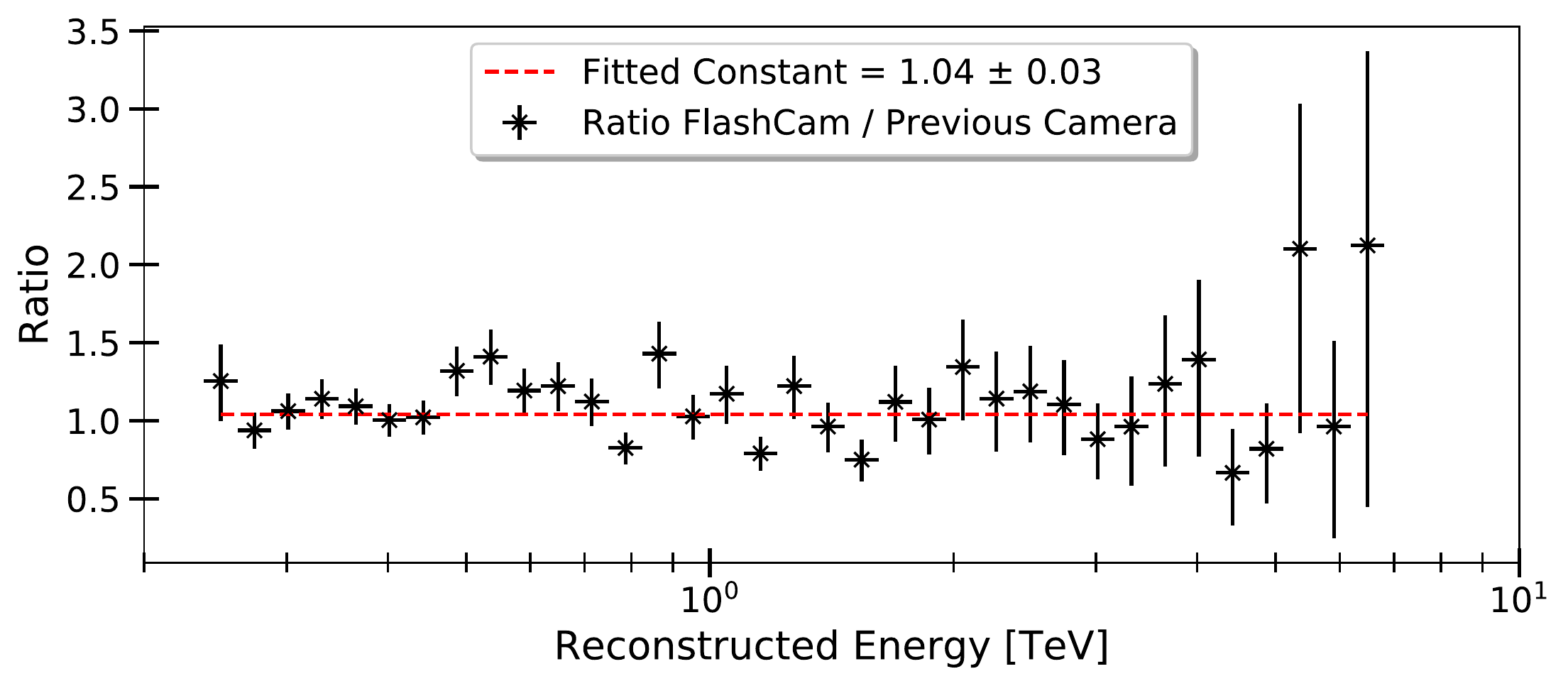} 
\end{center}
\vspace{-0.5cm}
\caption{A ratio of the Crab Nebula energy spectra derived with CT5-FlashCam and the previous camera, illustrating the compatibility of both spectra within statistical uncertainties.}
\label{fig:crabspectrumratio}
\end{figure}

As a representation of the verification of the full energy spectrum reconstruction with the reference source Crab Nebula, we show the ratio of the CT5-FlashCam spectrum and a spectrum derived with the previous camera from 2018 (Fig.\,\ref{fig:crabspectrumratio}). As expected, the ratio is flat, with a mean value that is within statistical errors compatible with 1.0. 
Nevertheless, some uncertainties of the calibration 
of these recent CT5 data sets are still under investigation at submission of this paper.

\begin{figure}
\begin{center}
\includegraphics[width=0.49\textwidth]{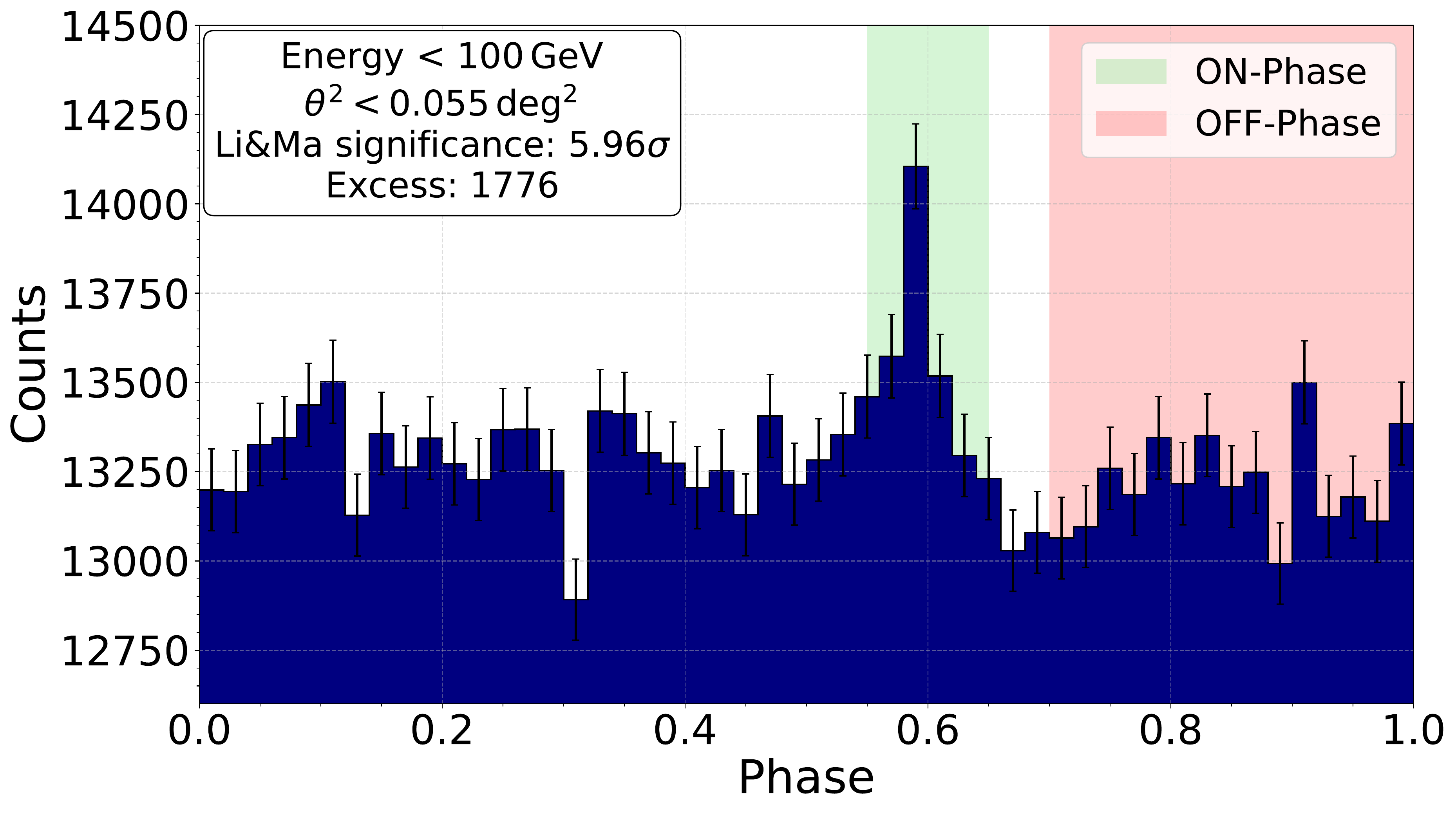}
\hfill
\includegraphics[width=0.49\textwidth]{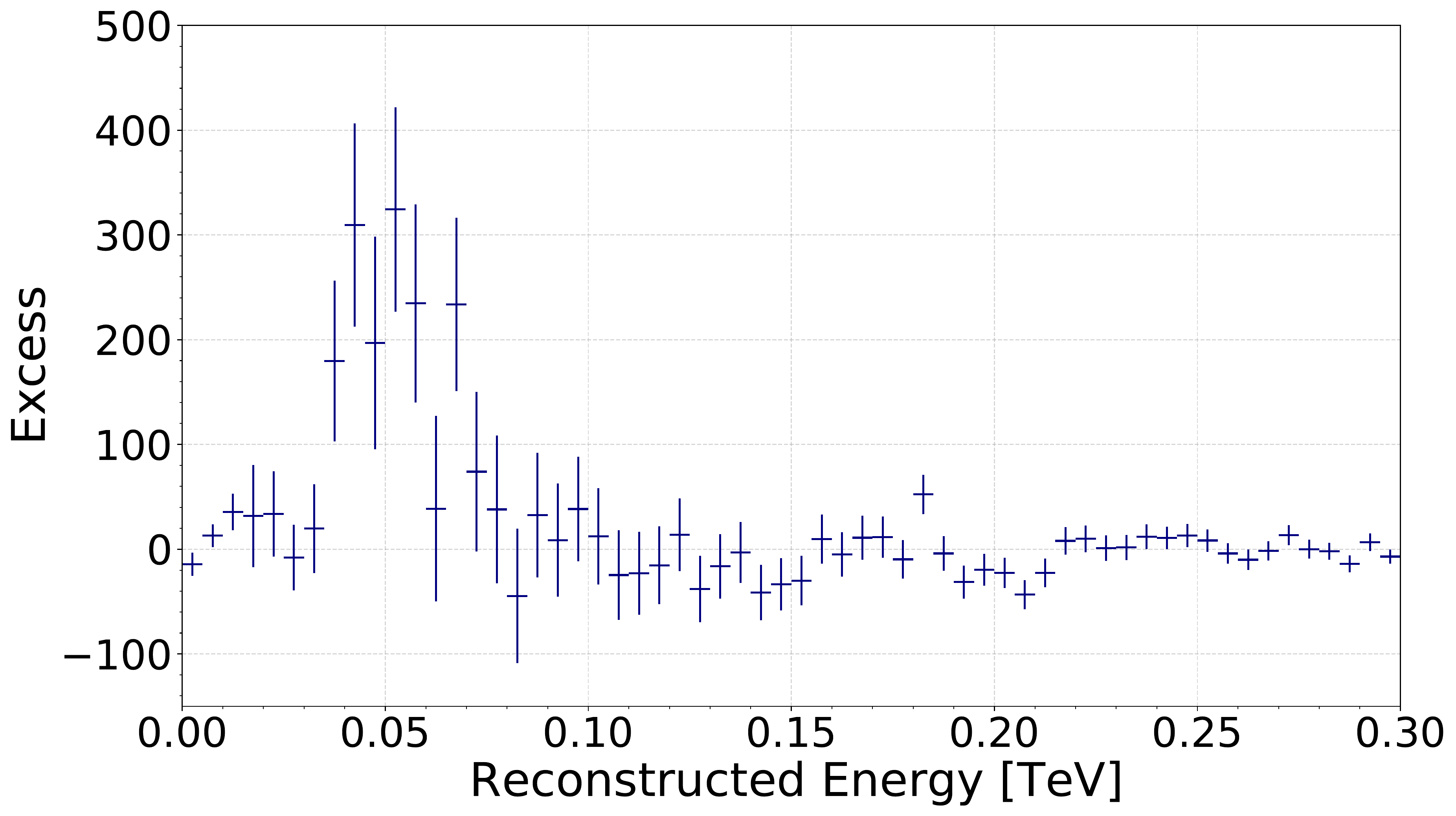}
\end{center}
\vspace{-0.5cm}
\caption{Left: phasogram of the Vela pulsar $\gamma$-ray data. Right: $\gamma$-ray energy distribution derived from the on-phase, after subtraction of the properly normalized off-phase distribution.}
\label{fig:vela}
\end{figure}

Finally, Fig.\,\ref{fig:vela} shows the phasogram of all $\gamma$-ray-like events below 100\,GeV derived from the Vela pulsar observations. Here, a loose cut configuration was chosen (image amplitude $> 25$\,p.e., angular distance cut $\vartheta^2 < 0.055\,\mathrm{deg}^2$), reflecting the very steep spectrum of the source ($\propto E^{-4}$ at $\sim$20\,GeV with an exponential cutoff, \cite{HESSVela2018}). The phasogram illustrates that the camera's event time stamping works as expected. Because of the steep spectrum, the reconstructed energies are a good measure of the energy threshold of this data set. The energy bias (derived from simulations) is of order 50\% at the peak of the reconstructed energy distribution of $\sim$$50$\,GeV, the peak of the true energy distribution is therefore estimated to be at $\sim$$35$\,GeV.

\section{Conclusions}

We have presented results from the science verification program after installation of a FlashCam (an advanced prototype of the camera model) into H.E.S.S.-CT5. No results from stereoscopic events with the other telescopes of the H.E.S.S.\ array were included at this time, to focus solely on CT5-FlashCam's performance. While the results show that the camera works up to expectations, optimizations of the data reconstruction algorithms are still ongoing, to fully exploit the scientific potential of the ensemble of the world's largest Cherenkov telescope and the first Cherenkov camera based on a fully-digital readout system.

\vspace{3ex}
\noindent 
\textbf{Acknowledgements:} 
H.E.S.S.\ collaboration acknowledgements can be found under this \href{https://www.mpi-hd.mpg.de/hfm/HESS/pages/publications/auxiliary/HESS-Acknowledgements-2021.html}{link}.
We thank Matthew Kerr and the \textit{Fermi}-LAT collaboration for providing the Vela pulsar ephemerides that were used in this paper.

\clearpage
\section*{Full Authors List: \Coll\ Collaboration}
%
\scriptsize
\noindent
H.~Abdalla$^{1}$, 
F.~Aharonian$^{2,3,4}$, 
F.~Ait~Benkhali$^{3}$, 
E.O.~Ang\"uner$^{5}$, 
C.~Arcaro$^{6}$, 
C.~Armand$^{7}$, 
T.~Armstrong$^{8}$, 
H.~Ashkar$^{9}$, 
M.~Backes$^{1,6}$, 
V.~Baghmanyan$^{10}$
% ...

\medskip

\noindent
$^{1}$University of Namibia, Department of Physics, Private Bag 13301, Windhoek 10005, Namibia\\
$^{2}$Dublin Institute for Advanced Studies, 31 Fitzwilliam Place, Dublin 2, Ireland\\
$^{3}$Max-Planck-Institut f\"ur Kernphysik, P.O. Box 103980, D 69029 Heidelberg, Germany\\
$^{4}$High Energy Astrophysics Laboratory, RAU,  123 Hovsep Emin St  Yerevan 0051, Armenia\\
$^{5}$Aix Marseille Universit\'e, CNRS/IN2P3, CPPM, Marseille, France\\
$^{6}$Centre for Space Research, North-West University, Potchefstroom 2520, South Africa\\
$^{7}$Laboratoire d'Annecy de Physique des Particules, Univ. Grenoble Alpes, Univ. Savoie Mont Blanc, CNRS, LAPP, 74000 Annecy, France\\
$^{8}$University of Oxford, Department of Physics, Denys Wilkinson Building, Keble Road, Oxford OX1 3RH, UK\\
$^{9}$IRFU, CEA, Universit\'e Paris-Saclay, F-91191 Gif-sur-Yvette, France\\
$^{10}$Instytut Fizyki J\c{a}drowej PAN, ul. Radzikowskiego 152, 31-342 Krak{\'o}w, Poland\\
% ...

\end{document}